\documentclass[12pt,draftclsnofoot,onecolumn]{IEEEtran}
\IEEEoverridecommandlockouts 
\ifCLASSINFOpdf
\usepackage[pdftex]{graphicx}
\graphicspath{{./Images/}}
\DeclareGraphicsExtensions{.pdf,.jpeg,.png}
\else
\fi

\usepackage[cmex10]{amsmath}
\usepackage{siunitx}
\usepackage{cite}
\usepackage{psfrag}
\usepackage[utf8]{inputenc}
\usepackage[T1]{fontenc}
\usepackage{amsmath,amsfonts,amsbsy,amssymb}
\usepackage{mathabx}
\usepackage{mathrsfs}
\usepackage[nolist]{acronym}
\usepackage{tabularx}
\usepackage{amssymb}
\usepackage{amsmath}
\usepackage{graphicx}
\usepackage{cite}
\usepackage{multirow}
\usepackage{wasysym}
\usepackage{multirow}
\usepackage{float}
\usepackage{color}
\usepackage{caption}
\usepackage{subcaption}

\begin{document}
\title{Why Smart Appliances May \\Result in a Stupid Energy Grid?}

\author{Pedro H. J. Nardelli and Florian Kühnlenz
\thanks{Pedro H. J. Nardelli is with Laboratory of Control Engineering and Digital Systems, School of Energy Systems, Lappeenranta University of Technology, Finland. Florian Kühnlenz is with the Centre for Wireless Communications (CWC) at University of Oulu, Finland. P.H.J. Nardelli is also with CWC. This work is partly funded by Finnish Academy  by Strategic Research Council/Aka BCDC Energy (n.292854). Contact: pedro.nardelli@lut.fi}%
}

\maketitle


\begin{abstract}
This article discusses unexpected consequences of idealistic conceptions about the modernization of power grids. It focuses on demand-response policies based on automatic decisions by smart home appliances. 
Following the usual approach, individual appliances sense a universal signal (namely, grid frequency or price) that reflects the system state.
Such information is the basis of their decisions. 
While each device has a negligible impact, their aggregate effect is expect to improve the system efficiency; this is the demand-response goal. 
The smartness of such an ideal system, composed by isolated appliances with their individual decisions, but connected in the same physical grid, may worsen the system stability. 
This undesirable outcome comes from the synchronization of the devices' reactions when subject to the same signal. 
We argue that this is a predictable effect of (implicit) methodological choices.
We employ a different approach that understands the electricity system as constituted by physical, informational and regulatory (networked and structured) layers that cannot be reduced to only one or two of them; it needs to be viewed as an organic whole so proper management tools can be designed. 
Two examples are provided to illustrate the strength of this modeling.
\end{abstract}

\section{Introduction}

The recent development of communication and information technologies have stimulated the energy utilities and electricity network operators to upgrade the long-time established power grid automatic control systems \cite{Bush2014b}.
Through this modernization path, more devices that react to information signals are appearing as part of the system physical structure. 
Among many other applications, these elements -- particularly the smart home appliances and smart meters -- are the technological bridge to involve the small-scale end-consumers as part of demand-side management policies \cite{Palensky2011}.
From this perspective, one can classify this modernization process as a socioechnical system \cite{Ilic2016}.
Following the hype, the operation and management of the modern grid tend to become more and more autonomous, given origin to the term smart electricity grid.
Such a system grows in complexity by having more elements, interactions and dependencies.
If this is the case, traditional reductionist methodologies usually lead to a poor understanding of the system-level dynamics and, even worse, misplaced interventions \cite{nardelli2014models,Bale2015}. 
As a contradictory unity, the system becomes more resilient and more fragile at the same time.

We identify here a mismatch between the (hidden) methodology applied to design autonomous appliances that shall be  deployed as part of the power grid.
Specifically, a kind of methodological individualism \cite{individualism}, as a mainstream approach, appears in this context as follows: A problem is delivered to an engineer, who solves it by developing elements that will react to a signal individually, considering that everything else is given.
In practice, however, other elements will likewise react to that signal (if designed to do so).
As they are interconnected through the grid, their aggregate action may lead to unexpected events from the methodological individualism point of view.


To move beyond this methodological weakness, we systematize a different methodology in \cite{Kuhnlenz2016} and other works thereafter.
The core idea is that sociotechnical systems shall be analyzed through three constituent layers: (\textit{i}) a physical layer composed by material things and connections, (\textit{ii}) an informational layer related to symbolic classifications and communications, and (\textit{iii}) a regulatory layer involving decision-making procedures and rules. 
The relation from the physical to the informational layer happens through sensing.
The information sensed and processed by the agents is the basis of their individual decisions, which may or may not be the same.
These decisions may then result into actions that may affect the physical and information layers, directly or indirectly.
The system is then constituted of (and not reduced to) these layers.

To illustrate such an approach, we have developed two computational study-cases that present problems arisen from the methodological individualism.
One case is inspired by the frequency stabilization via smart fridges, as introduced by \cite{Evora2015}.
{In this contribution, the authors showed different methods to coordinate the fridges' reactions via randomization, similar to random access control in communication networks. For example, if the frequency drop is perceived by a fridge, it waits a random time to become activated (i.e. using energy from the grid), and also to return to its normal cycle operation after the situation is resolved. This policy serves to avoid the oscillations from the synchronized collective reactions.}

The other case is inspired by \cite{Krause2013}, where the authors present an analysis of how smart washing machines may answer to dynamic price schemes.
{In their own words: ``(...) when agents are exposed to source noise via correlated price  fluctuations (as adaptive pricing schemes suggest), the market may amplify those fluctuations. In particular, small price changes may translate to large load fluctuations through catastrophic consumer synchronization. As a result, an adaptive power market may cause the opposite effect than intended: Power demand fluctuations are not dampened but amplified instead.''}
For both cases, the expected improvement in the system stability falls short, producing unforeseeable, more unstable, dynamics.
This emergent behavior is due to the synchronization effect induced by the collective reactions of smart appliances.
This collective behavior -- blind for individualist methodologies -- creates a dynamic sequence of overshoots and/or undershoots.
Our results, in consonance with \cite{Evora2015,Krause2013}, show that these events may be controlled, avoided or mitigated by interventions designed through the proposed three-layer approach.

\subsection*{Collective behavior}
{Collective behavior as an emergent phenomenon has been extensively studied in complex system sciences \cite{Newman2011}; the topics are somehow universal, ranging from  markets and ecosystems to ``the whole of human society.''
Looking at technological solutions, swarm intelligence is probably the most known way of using the emergent features of collective behavior to solve computational problems in different fields (refer, for instance, to \cite{Blum2008}).
In \cite{Evora2015}, Evora et al. provide an in-depth review of swarm intelligence and the advantages to apply it in smart grids by creating a more resilient system.}

{Market institutional arrangements, in their turn, are characterized by signaling demand-supply relations via price.
It can be then modeled as composed by individual entities that ``answer'' to such signals (e.g. buying or not buying, or selling or not selling).
Among different currents, the growing field of Complexity Economics (e.g. \cite{Farmer2009}) tries to cope with theoretical failures that neoclassical standard economical approaches cannot handle.
By doing so, they open similar methodological questions as the present article.
The paper by Krause et al. \cite{Krause2013} follows this line-of-thought by showing how a naive market model may result in a ``catastrophic'' outcome to the power system.}

\section{Coordination through signals}
Coordinating individual agents using signals is widespread and proves its effectiveness every day.
Traffic lights are a good example: drivers stop when the light is red and move when is green.
Other example: some health centers where the patient upon entry selects one from different options related to his/her health condition to then get a number assigned.
Different numbers appear in screens at waiting rooms, each number associated to the next person to be served.
In both cases, an internalized rule-based coordination exists.
Imagine possible situations without it: messy traffic, crazy pushing and pulling in lines, discussions about treatment priorities etc. 

Nevertheless, although both situations indicated a need for coordination, their specifics and the respective problems to be solved are quite different.
So are the correspondent solutions.
But, how could one assess these differences?
Let us analyze the traffic light example.
When a driver sees the red light, he/she stops not only because it is written in the norm, but also because he/she knows that all others involved in the traffic know this.
Every driver expects that all others are going to behave based on the signal given by the light in the same way that he/she would.
Due to the material characteristic of the situation (two things cannot occupy the same place and car accidents are undesirable), the coordination policy is almost always effective. 

Turning our attention to the health center example, the person knows his/her number, but it is usually hard to know how many people are in his/her line or even how many lines exist.
So it is hard to estimate how long it is going to take and if the situation is fair based on his/her own health state compared to the others. 
The  coordination mechanism structure and its effectiveness may be unclear to the persons involved (although it may be clear to the designers).

From these two every-day examples, we suggest three different ways of classifying the system, following three layers of analysis \cite{Kuhnlenz2016}:
\begin{enumerate}
	\item \textbf{Physical layer (PHY)} relates  to the material problem to be solved -- who can move the car or who is served by the nurse/physician. The relations between the elements are normally related to transportation (e.g. flow of cars or people or electrons). In this case, physical laws determine the dynamics.
	\item \textbf{Information layer (INF)} relates to the access to information -- ``who knows what?'' In this case, information can be obtained through (a) sensing and processing data from PHY, or (b) through communication between agents in INF, or (c) through total or partial broadcast from agents in REG (to be defined next).
	\item \textbf{Regulatory layer (REG)} relates to the regulatory action -- how the individual decision is made and how is the (re)action to it. The objective of the agent can be anything from maximizing their individual pleasure to following a simple threshold rule or even random choices. The decisions may render changes in PHY by acting upon physical devices (e.g. moving the car) or in INF by broadcasting information to one, or to a group, or to all other agents.
\end{enumerate}


This article is built upon the understanding that all three ways are equally important; they reflect three different layers where events happen.
They constitute and structure the sociotechnical system under analysis.
As constituent layers, the system, in general, cannot be reduced to any of them, otherwise the results and policy guidelines may lead to undesirable outcomes.
We will show along the rest of this article how the methodological reductionist bias may be harmful when smart appliances are employed as demand-response tools designed to help the operation of modern electricity power grids.

\section{Demand-response and smart appliances}

Smart home appliances -- as part of the Internet of Things (IoT) trend --  are expected to play a big role in the modernization process of the electricity grid. %
With the increase of intermittent sources of energy (e.g. solar and wind) as substitute of more controllable ones (e.g. thermal), electricity supply becomes less predictable than it is now.
Then, demand is expect to respond accordingly, employing flexibility in consumption instead of production.
Smart appliances are planned to react by adjusting their usage according to the available power.
In this case, two universal coordination signals appear: frequency and price.

\subsection*{Frequency}

In electricity grids, the frequency of the alternate current shall be constant, reflecting a balanced supply and demand.
In Europe, the nominal frequency is 50 Hz. 
If the supply is higher than demand, the frequency rises. 
On the other hand, if supply is lower than demand, frequency drops. 
This effect is a fundamental part of the physics of the current system.
The grid frequency is the same for regions connected in the same power grid (synchronous region).
In other words, any appliance connected within a given synchronous grid experiences the same frequency at the same time.
As frequency is an indicator of supply and demand balance in real-time, it can be used by smart appliances as a signal to guide demand-response.
Fridges, for instance, could adjust their cooling cycles as a reaction to frequency deviations \cite{Evora2015}.
If its value is too low, based on a given threshold, the smart appliance postpones its cycles to reduce the load on the system, and vice-versa.

\subsection*{Price}

Dynamic ``real-time'' electricity price can also be seen as a universal indicator of the supply and demand situation.
Price is a metric that reflects (at least in theory) the willingness or need of buying by the consumers and availability of supply.
In the electricity wholesale markets, it indicates the most expensive power produced to match the demand in a given period of time \cite{Kuhnlenz2017}.
It is worth noting that price and frequency are associated to different timescales.
While frequency is a direct measure of the physical grid (PHY), it is fair to say that it captures the system state in real-time.
The real-time in price, on the other hand, is different; it is only an indirect measure of the physical reality, which must be related to some period of time, generally one hour.
In this case, price is a construction in REG based on predictions of all supply and demand of that specific period.
In an electricity system dominated by intermittent generation from solar and wind, for instance, one expects fluctuations in supply.
Then, smart appliances that have more flexibility in their usage (e.g. washing machine) can wait until more power is available -- reflected by lower  electricity prices -- to be turned on.
In this way, the situation becomes win-win: the consumer pays less and the system become more balanced \cite{Shariatzadeh2015}.

\section{Synchronization and system-level coordination}

In both cases discussed in the previous section, the appliances are aware of the same signal, either frequency or price, depending on the timescale to be considered. 
However, if the appliances' decision procedures are based on the individual behavior without considering how the others may react to the signal, possible undesirable fluctuations may emerge (even when the goal of stability is shared by all individual agents).
In other words, this kind of methodological individualism -- widespread in many fields -- may lead to optimal individual solutions when the others are assumed external elements, but this may lead to poor solution for the whole system \cite{nardelli2014stable}.

\begin{figure}
	\centering
	\includegraphics[width=0.75\columnwidth]{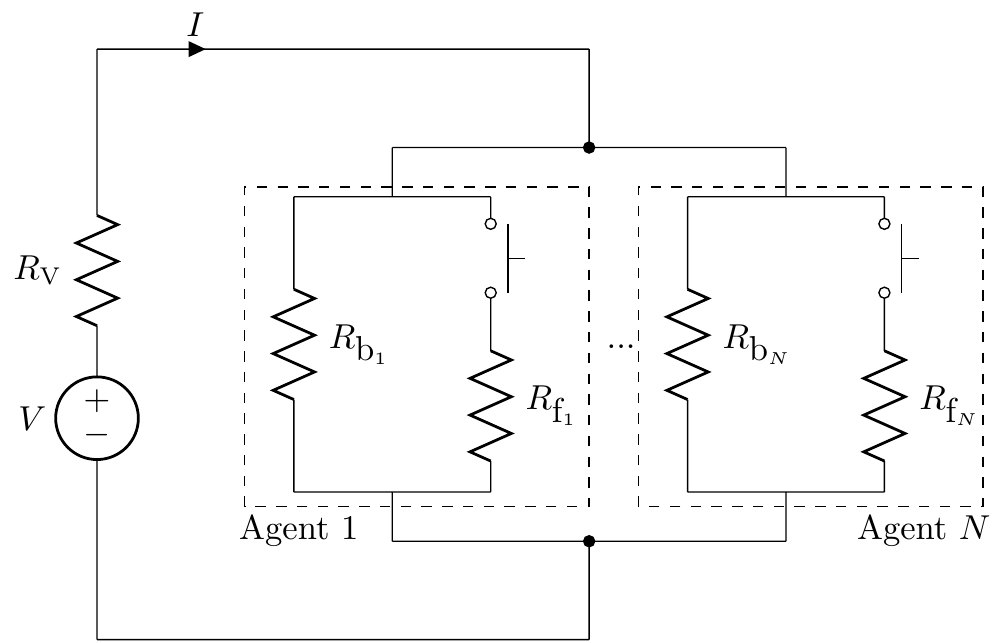}
	\caption{Electrical circuit representing the physical layer of the system. The circuit is composed by a power source $V$ and its associate resistor $R_\textup{V}$, and resistors of in parallel, generating a current $I$. These resistors are related to $N$ agents, where each agent $i=1,...,N$ has a basic load $R_{\textup{b}_i}$ and controls a flexible load $R_{\textup{f}_i}$.}
	\label{fig:System} 
\end{figure}

To investigate these scenarios, we employ a variation of the models presented in \cite{Evora2015,Krause2013} following a modified version of the multi-layer system introduced in our previous work \cite{Kuhnlenz2016}.
Our discrete-time agent-based model assumes an electric circuit as the physical infrastructure as illustrated in Fig. \ref{fig:System}.
Although this model is based on direct current, clearly a very simple one, it captures the essential features of the system to be investigated. 
In this case, instead of frequency, the stability of the system is evaluated by the voltage experienced by the agents, which reflects the physical balance between supply and demand.
The documentation and source-codes of the proposed experiments can be found at  \cite{github-flo}.

\subsection{Direct voltage control}
This experiment is a simplified DC version of the one proposed in \cite{Evora2015}, which focused on frequency control in alternate current scenarios.
We start assuming each individual smart appliance acts based on the voltage experienced by every agent.
If the voltage is below a given predetermined threshold, the appliance postpones its planned activation (emulating fridge cycles).
Using this policy, a coordination will happen: every smart appliance will act synchronously. 
This phenomenon -- probably unexpected in methodological individualist approaches -- yields  overshoots and undershoots that, instead of stabilizing the voltage (or frequency), create a system that is structurally unstable.

\begin{figure*}
	\centering
	\includegraphics[width=1\textwidth]{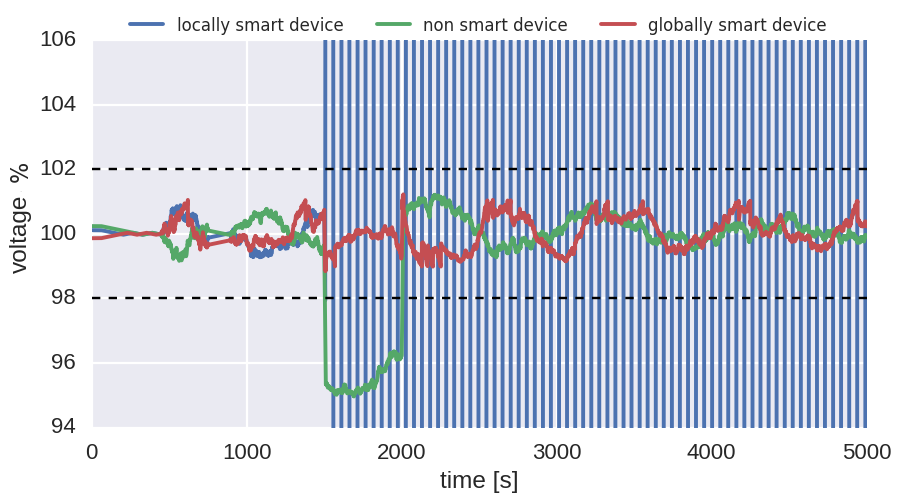}
	\caption{Simulation comparing the voltage control for three scenarios. The green line refers to the basic scenario without voltage control. The blue refers to the smart control with devices that decides locally (individualist approach). The red refers to the proposed solution that considers the global behavior of the system. \label{fig:simulation2}}
\end{figure*}

{Fig. \ref{fig:simulation2}  shows an example of such dynamics.
Note that the green line is the reference voltage, which has a sudden drop below the 98\% threshold at time 1500. 
The blue line represents the voltage behavior when all devices are individualistically reactive to the signal.
One can observe that, after the drop, a voltage spike happens due to the synchronized reaction of appliances: they have postponed their cycles in the same way.
But, such a spike drove the system out of its desired operation so that the appliances need to react again since there is oversupply in the system.
The collective reaction leads to another drop, but much worse than the initial one. 
Even more dramatic, the system dynamics become oscillatory, even after the initial voltage drop is restored.}

Note that the issue is systemic from the way that the appliance reaction was designed.
More interestingly, this happens even when all fridges have the same shared goal of stabilizing the grid.
The problem is then methodological in its core, and not result of a bad technology.
The smart appliances are doing what they are supposed to do, so they are still smart in this sense.
However, the system is more unstable, going against the final shared objective.
The system and its elements are indeed reactive to voltage variations above/below the established limits, but not in a smart way.
And yet, the problem cannot be reduced to the single agents alone: the issue is a structural one and happens due to the way the agents are organized/coordinated.

\begin{table}[!t]
	\centering
	\caption{Three-layer description}
	\label{tab}	
	\centering
		\begin{tabular}{|p{1.2cm}|p{4.5cm}|p{4.5cm}|p{4.5cm}|}
			\hline
			\textbf{Scenario}	&	\textbf{PHY} &	\textbf{INF} &	\textbf{REG}	\\ \hline \hline
			\multirow{2}{*}{Direct} 
					& Structure given in Fig. \ref{fig:System} \newline Goal: Stable voltage value	 & Voltage information from PHY	\newline Knowledge about others & Action: Add or remove load \newline Goal: Stabilize voltage  	\\ \hline
			
			\multirow{3}{*}{Price} 
					& Structure given in Fig. \ref{fig:System} \newline Goal: Stable voltage value & Price information from REG \newline Voltage information from PHY	 \newline Knowledge about others &	Action: Add or remove load \newline Goal: Minimize the cost\\ \hline
			
		\end{tabular}
\end{table}

Let us now consider that the appliances are designed to consider that the
other appliances with the same design are also connected in the physical grid.
Our three-layer approach now becomes an important tool for building an effective policy towards the system-level goal, as presented in Table \ref{tab}.
The system goal given in PHY and the individual appliance goal in REG are the same: stabilizing the voltage when  instabilities appear.
From the information layer, the appliances have direct access to the system state.

As discussed before, the problem appears because all appliances react synchronously.
So the solution to the problem is coordinate the appliances' reactions, similar to medium access control techniques used communication networks \cite{card2011}.
A central controller can just inform which appliance is to be turned on in a given period (time-division approach).
A decentralized approach -- the one used here -- is based on randomization: if an appliance experiences a situation that action is needed, then it will only act with a given probability or after a random period of time (like Aloha or carrier sensing protocols).
The red curve in Fig . \ref{fig:simulation2} illustrates the behavior of the proposed solution, showing that (if the system parameters are properly tuned) the goal of stabilizing the voltage can be achieved.

\subsection{Voltage control via dynamic pricing}
\begin{figure*}
	\centering
	\includegraphics[width=1\columnwidth]{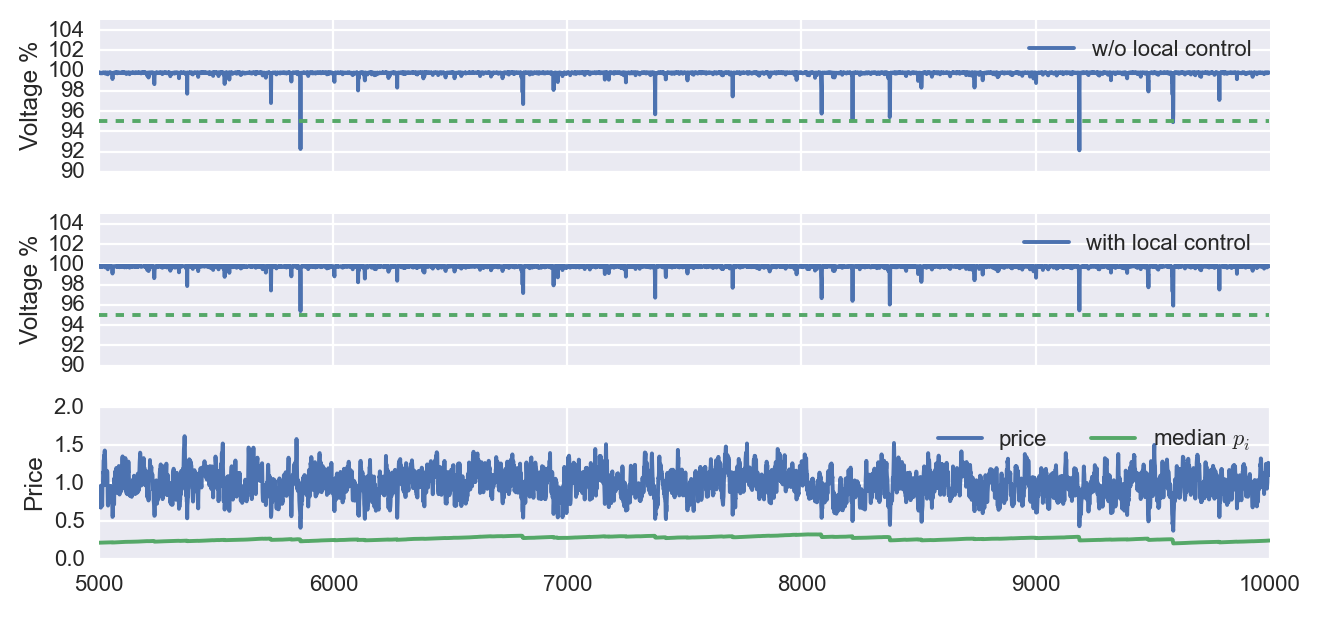}
	\caption{Two simulation runs showing the differences between the scenarios without (top) and with local voltage check (middle). The bottom plot depicts the market price. The dashed green line represents 5\% drop, which indicates the limit. \label{fig:simulation}}
\end{figure*}

We propose a similar experiment (although in different timescales) based on smart washing machines that react to dynamic price signals.
Consistent with \cite{Krause2013}, this scheme may worsen the system stability in terms of voltage variations instead of softening it.
Peaks in demand will occur as a collective, synchronous, reaction to low prices.
These spikes in electricity demand may, and probably will, be harmful to the power grid.
The predicted win-win situation turns out to be idealistic.
The top and the bottom plots in Fig. \ref{fig:simulation} show outcomes of this scenario using our proposed model.
Following \cite{Krause2013}, the sudden variation in the aggregate demand is caused by the internal agent state; these internal price expectations slowly synchronize during high price periods.
The synchronized collective reaction is then triggered when a lower price appears.

\begin{figure}
	\centering
	\includegraphics[width=0.75\columnwidth]{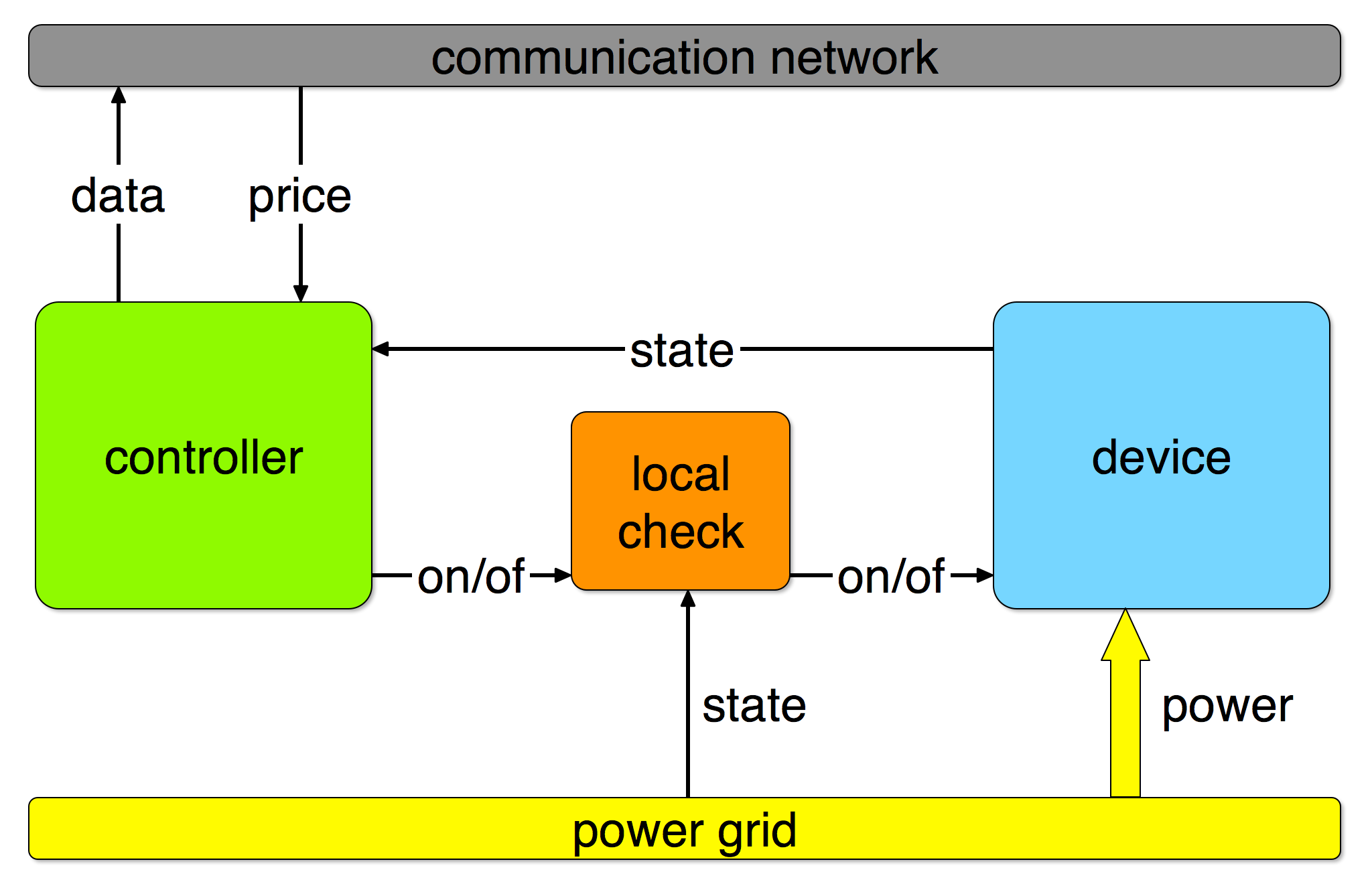}
	\caption{Diagram of the proposed solution. The smart meter is not in direct control of loads, but rather the signal has to go through a voltage-check layer that will be used to decide about the load usage. \label{fig:system}}
\end{figure}

As argued before, this is not an unfortunate event: it is rather a systemic feature from the methodological individualism embedded in the demand-response policy design and the respective individual appliances' reactions.
The proposed three-layer approach can be used to mitigate the issue by building interventions that make explicit the structure of the phenomenon, as presented in Table \ref{tab}.
A schematic of our proposed solution is presented in Fig. \ref{fig:system}, where the decision procedure of individual agents  is based on the physical network state and the price associated to that period.
Fig. \ref{fig:simulation} presents an output of a simulation that compares the system behavior with (middle plot) and without (top plot) the local voltage check for the same starting conditions and same price curve.
The local voltage check strategy leads to much lower drops as far as the agents look not only to the price to make their decision, but also consider what is actually happening in the physical system.
In other words, they do a ``sanity-check'' to verify whether the price is really giving the most appropriate signal based on the voltage level observed before the decision of turning on the flexible load. 

\section{Final remarks}

Although our simulation results were obtained through a simple model, they are inspired by material, real-world, phenomena.
In other words, our approach tries to get the essence of the electricity system -- defined by the electricity interchange operation -- by analyzing the concrete dynamics of system in action.
Our approach acknowledges that the individual behavior is part of a complex system that has different structures in different layers.
We argue that this consideration is a necessary condition to have an effective demand-response policy.

Smart appliances that are designed to work selfishly can be deployed in the grid in small number up to a certain point.
If their usage scales up (as it is claimed and expected), then the concerns posed by this article become very relevant.
For instance, when 1\% of appliances avoid the peak evening hour due to high prices, they can indeed reduce peak-load and then decrease the usage of reserve power plants.
If this number grows to 10\%, they might create another unforeseen peak after the high price hours, leading to under-usage during the high price hours and reserve power needs in the low price hours afterwards.

\begin{figure*}
	\centering
	\includegraphics[width=1\textwidth]{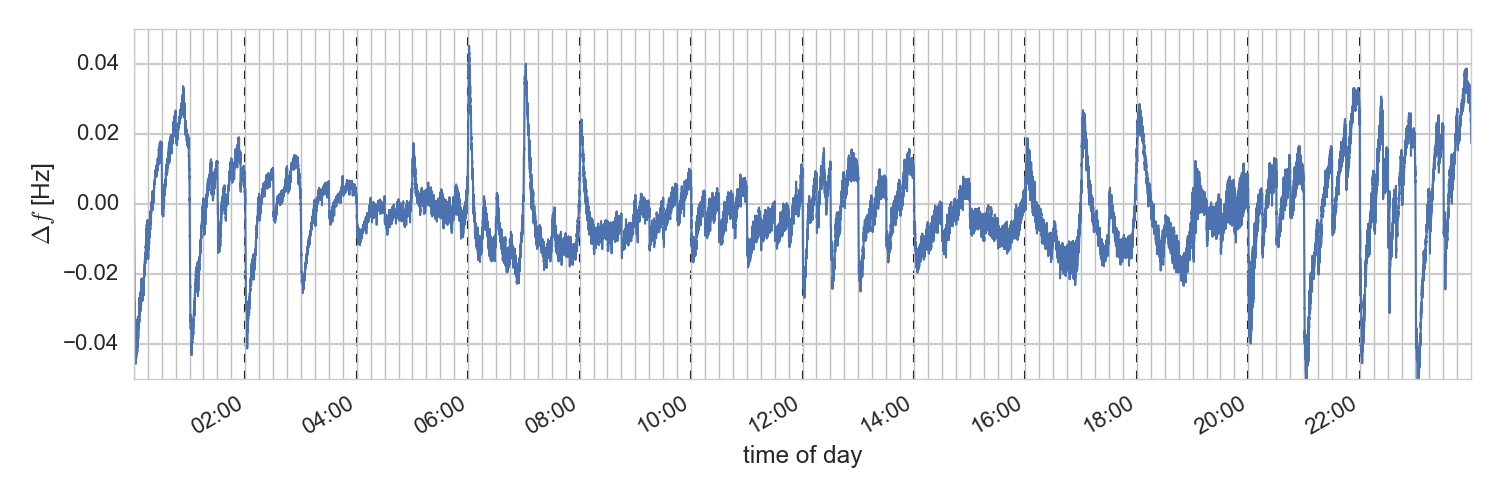}
	\caption{Average grid frequency over more than one year. The effects of the market timing, with a bidding period ending every 15 minutes, are clearly visible. This illustrates the coupling between markets and the physical system. Data is available in \cite{netzsin}. \label{fig:frequency_average}}
\end{figure*}

This might seem unlike at first, but even nowadays it is possible to see the structural effects of the wholesale electricity market in the physical grid.
Fig. \ref{fig:frequency_average} presents the average grid frequency for every second of the day for a period of over twelve (12) months in Germany.
As the nominal frequency value in Europe is 50 Hz, one would expect the average would converge to such value, regardless of the measurement time. 
However, one sees a quite different behavior, where two patterns can be identified.
The first one being that more intense spikes occur every full hour while, smaller ones occur every quarter hour; these specific spots are the time-ticks of the European electricity market EPEXSPOT \cite{epex}.
The other one is that, for some hours, we see negative deviations while in other hours we see positive ones.
This suggests daily reoccurring over/under corrections.
In fact, a very recent paper discusses in-depth this specific case \cite{Schafer2018}.

Likewise, there was a problem in Germany related to the upper limit for normal operation of the grid frequency, as discussed in \cite{50.2hz}.
This value is set to 50.2 Hz.
For this reason, the initial legislation demanded that all photo-voltaic generation must shut down when that value was sensed. 
With the high penetration of such a source, the aggregate effect of this rule-based behavior had the potential to decrease the power generation up to 9 GW (approximately the capacity of ten big thermal power plants).
Such an abrupt generation loss would lead to a cascade effect, causing a blackout in Europe.

{Moreover, the predicted increase of storage and electric vehicles introduces another challenge to the grid management.
Scaling-up the use of storage  may create problems in the operational timescale (like the proposed smart fridges)  and the market timescale (like the proposed washing machines).
In the first case, batteries may synchronize their activation/deactivation cycles in relation to the frequency (voltage) signal.
In the second case, they may be used to speculate in the market: buying when is cheap, selling when is expensive. 
This may help the system stability, but also may create synchronization of expectations (like the washing machines) that harms its operation.
Clearly, the proposed models cannot be extended to such cases, but it may indicate another important research direction; if distributed storage capabilities are to scale up, they need to be properly organized and coordinate to achieve the desired positive effects in the grid (refer to \cite{Auer2017} and references therein).}

All these undesirable, emergent, phenomena are due to synchronization of individual actions and reactions.
As we have argued here, this is a structural feature that emerges from the system design that (unconsciously) assumes a methodological individualism approach as default. 
Proper interventions shall be based on models where the complexity of interactions across and along physical, information and regulatory domains are understood as constituent parts of the system.
Otherwise, there is a big risk of smart parts build a stupid whole due to structural reasons hidden in methodological choices.

\vspace{-1ex}
\bibliographystyle{IEEEtran}

\end{document}